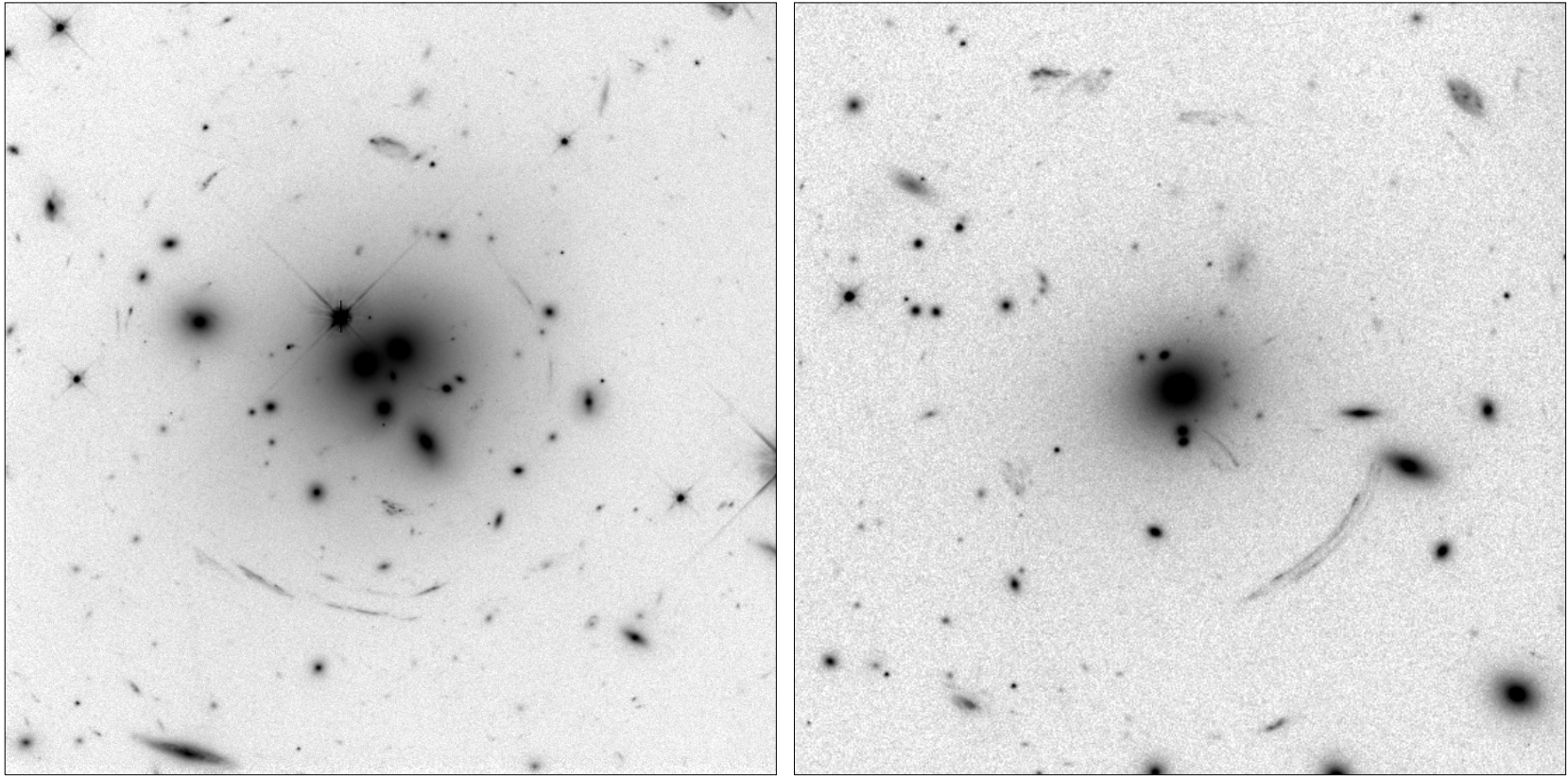

FIGURE 1. HST images of two EMSS clusters with giant arcs: MS 0440+0204 ($z$=0.19) at the left and MS 2137−2353 ($z$=0.31) at the right. Both images are 10-orbit exposures in the F702W filter. Note the thin, radial structure in MS 2137−23, first discovered by Fort et al. (1992) and later modelled as a radial gravitational image from a cluster with a finite core by Mellier et al. (1993).

# X-ray Selected Clusters of Galaxies

Isabella M. Gioia[1]

*Institute for Astronomy, University of Hawai'i , Honolulu, HI 96822*

*Istituto di Radioastronomia del CNR, 40129 Bologna, Italy*

**Abstract.**

Three topics will be presented in this paper: 1) study of the clusters and groups of galaxies found serendipitously in the North Ecliptic Pole (NEP) region of the ROSAT all-sky survey; 2) the highest redshift clusters found in the EMSS and the cosmological implications of their very existence; 3) the gravitational lensing in the EMSS X-ray selected clusters of galaxies observed by the Hubble Space Telescope.

## 1. Introduction

The study of clusters of galaxies especially at high redshift is very important since it yields information relevant to cosmology. Clusters of galaxies are the largest bound objects in the Universe, thus they provide constraints on the formation of structure and on the composition of the Universe. Clusters are also very luminous X-ray emitters, hence they are found in X-ray surveys of the sky. Selection of clusters, particularly those at high redshift, by means of their X-ray emission is one of the cleanest ways to avoid sample contaminations. However, one has to follow up with spectroscopic observations to obtain information useful for cosmological studies. All the information provided by clusters is only as reliable as the catalogs they come from. Biased catalogs give inevitably biased results. In the past, galaxy clusters have been selected from optical images using subjective selection criteria, often applied to inhomogeneous plate material (see for instance Abell et al. 1989; Gunn, Hoessel and Oke 1986; Couch 1991). Even the compilers of these catalogs have cautioned against using them for statistical purposes. A distinct improvement over the previous mentioned catalogs is constituted by the use of computer algorithms to select clusters from digitized plates (see among others Dalton et al. 1994, etc.). However, first, clusters are chosen in the absence of redshifts and, second, fluctuations in the background galaxy density will cause: 1) inclusion of non existing clusters, 2) exclusion of real clusters. The present status is that there are several hundred clusters known with redshift z < 0.2, but less than one hundred clusters with measured redshift

---

z > 0.3 after decades of effort. None of the clusters at high redshift come from samples that are complete.

## 2. Clusters in the EMSS and ROSAT NEP surveys

Clusters are very luminous X-ray sources comparable to quasars. Identifying clusters through their X-ray emission circumvents most of the problems of optical selection. The X-ray selection is based on objective criteria which use computer-based algorithms, which are well defined and quantifiable and thus biases can be removed. Projection effects are much less severe than with optical selection, even if they may exist.

One X-ray selected cluster catalog, the EMSS cluster catalog by Gioia & Luppino 1994, is already available. It contains over one hundred clusters identified in the course of several years by the EMSS team (Gioia et al. 1990a, Stocke et al. 1992). Twenty-six of these clusters are at a redshift greater than 0.3 and six at a redshift greater than 0.55 (of which the highest is at 0.83, the most distant X-ray selected cluster!). The EMSS cluster catalog is probably the only catalog containing high z clusters, at present, which has been constructed with objective and quantifiable selection criteria. Several statistical studies have been performed using the EMSS catalog:

1. X-ray evolution of the luminosity function (Gioia et al. 1990b, Henry et al. 1992) indicates cluster evolution has occurred quite recently and the most luminous clusters have evolved the most. Clusters at $z \sim$ 0.6-0.8 are probably dynamically young since the sound crossing time of the intra-cluster medium in a typical cluster is a third of the Hubble time;

2. Velocity dispersion studies by the Canadian Network for Observational Cosmology (CNOC) group (Carlberg et al. 1994) have been performed to obtain estimates of cluster masses and measure the contribution to the closure density;

3. A search for lensed galaxies in the form of both giant arcs and arclets have been carried out by Luppino, Gioia and collaborators (see Luppino et al. 1993; Gioia et al. 1994; Le Fèvre et al. 1994);

4. An ongoing collaboration with Nick Kaiser is discovering weak lensing in EMSS clusters (MS1224+24 at z=0.33, Fahlam et al. 1994; MS1054−03 at z=0.83, Luppino and Kaiser, in preparation) The weak lensing is essentially the distortion of faint galaxies by the gravitational lens effect and can yield a direct measure of the 2-D projected mass density with no assumption about the dynamical state of the cluster.

In collaboration with Pat Henry at the University of Hawai'i (UH) we are extending the pilot work on the EMSS clusters with data from the ROSAT All Sky Survey centered on the NEP. We have access at Mauna Kea Observatory (MKO) to the large amount of optical telescope time required to separate out the clusters of galaxies from the other X-ray sources in the ROSAT data. We are collaborating with colleagues at Durham and Cambridge (England), and at SAO and STScI, who also have access to additional telescopes around the world.

The NEP survey is an improvement on the EMSS since the X-ray data are deeper and come from a contiguous region thus allowing in the future large scale structure investigations. There are 629 sources in an area of about 84 square degrees. For each source we have a position, an X-ray count rate and the X-ray iso-contour map. To accelerate the identification procedure and start using the data for studies of clusters and groups of galaxies, we have explored four regions of the identification parameter space: 1) observe sources at blank fields on Schmidt plates (possibly distant clusters); 2) identify completely two 2.5×2.5 square degree regions, each one with an average NEP exposure, to start building statistically complete samples; 3) study extended sources associated with bright galaxies in groups; 4) identify completely an extremely deep pointed observation centered on the NEP to faint flux level (Bower et al. 1995). The identification rate of the NEP is today about 35%. Comparing the different classes of astronomical objects (AGN, clusters, BL Lacs, etc.) from the two completely identified regions in the NEP with the EMSS we find similar percentages, except that we are expecting a total number of distant clusters in the NEP about a factor of 3 higher than known today. Results already obtained from the NEP are: the discovery of the most distant X-ray selected QSO (z=4.3, Henry et al. 1994); a very distant X-ray selected cluster at z=0.82, second only in redshift to the EMSS cluster MS1054−03 at 0.83; a large gravitational arc in the NEP cluster associated with A2280 (Gioia et al. 1995); determination of the ensemble X-ray properties of groups of galaxies (Henry et al. 1995). From the NEP we have extracted the first X-ray selected, statistically complete, though small, sample of groups of galaxies, only eight groups with z≤0.04. We have determined from these data the group X-ray luminosity, temperature and group mass functions (see Henry et al. 1995 for a detailed discussion of the results obtained).

## 3. The EMSS High-z Clusters

The access to MKO telescopes allows us to conduct optical studies of the EMSS distant clusters of galaxies. The following work and results have been obtained in collaboration with Gerry Luppino at the University of Hawai'i. We have taken deep exposures at the UH 2.2m telescope and determined additional spectroscopic redshifts using the Multi Object Spectrograph at CFHT 3.6m and the Low Resolution Imaging Spectrograph at the Keck 10m telescope. There are six X-ray luminous clusters (all with $L_x > 10^{44}$ erg s$^{-1}$) at redshift exceeding 0.55. They are all apparently massive. Two of them, MS2053−04 and MS0451−03 exhibit gravitationally lensed arcs. The most distant cluster in the EMSS, MS1054-03 (z=0.83) has an optical richness comparable to Abell richness class 4. Luppino and Kaiser (in preparation) report the detection of the gravitationally-lensed distortion of faint, distant background galaxies by this rich and X-ray luminous ($L_x$=9.3×10$^{44}$ erg s$^{-1}$) cluster, the first measurement of weak lensing at such high redshift!

The very existence of massive clusters at high redshifts is problematic for standard CDM theories of hierarchical structure formation. Evrard (1989) argued that the existence of only 3 clusters with high velocity dispersion in the intermediate range z ∼ 0.4-0.6 is difficult to reconcile with a high value of the bias parameter b (the ratio of the galaxy to mass fluctuations within a 8 h$^{-1}$

Mpc radius sphere). But these clusters do exist! We have six of them with z>0.55 from the EMSS and probably even more from the NEP survey (and at redshift as high as 0.8!). Luppino & Gioia (1995) have discussed the cosmological implications of the existence of these same clusters for structure formation theories such as $\Omega = 1$ CDM, hybrid $\Omega = 1$ C + HDM, and flat, low density $\Lambda$+ CDM models.

## 4. Arcs in X-ray Selected Clusters of Galaxies and HST Observations

In the course of the last three years we have conducted an observational program at UH to search for arcs and arclets in X-ray selected clusters (in collaboration with G. Luppino, O. Le Fèvre, F. Hammer and J. Annis). We assume that the X-ray gas is in hydrostatic equilibrium with the gravitational potential and thus serves as a tracer of the total potential caused by both the luminous and dark matter. By choosing clusters with large X-ray luminosities we select deep potential wells and thus true clusters, more likely to exhibit the lensing phenomenon. We have had a high success rate (see Fig. 1 in Gioia et al. 1994) for the detection of arcs using a complete sample of 40 EMSS clusters. The description of the sample is given in Luppino et al. (1993) which first reported the discovery of the spectacular lens system in MS0440+0204. Preliminary arc survey results are given in Le Fèvre et al. (1994).

Two of the EMSS clusters with gravitational lenses were observed by the WFPC2 with a 10-orbit exposure. Fig. 1 presents these two spectacular examples of clusters and shows the capabilities of the refurbished telescope. More than 14 arcs are found in the core of MS0440+02. In many aspects this rich cluster appears to be an ideal lens. Preliminary source reconstruction techniques (Hammer et al. 1995) reveal the internal structure of background source galaxies, which show arm-like structures and compact sources. These faint distant galaxies would be barely resolved from the ground without these HST high quality data. In the second cluster observed by HST, MS2137-23, the thin radial feature first discovered by Fort et al. (1992) and modelled as a radial gravitational image from a cluster with a finite core by Mellier et al. (1993) is definitively resolved. Please notice also the double structure of the tangential arc which was not visible from the ground.

## 5. Conclusions

The ROSAT NEP survey will provide a cluster catalog deeper than the EMSS and thus of great interest for evolutionary studies. We have measured for the first time the ensemble properties of X-ray selected groups from the NEP. The EMSS is providing massive and distant clusters which can be used to measure both the X-ray luminosity evolution and to test for weak lensing at higher z. X-ray selected samples have also been demonstrated to be an efficient way to find clusters at z as high as 0.8 and gravitational arc systems in clusters at z as high as 0.6.

**Acknowledgments.** I am grateful to all the collaborators involved in the various projects who allowed me to discuss these results: P. Henry, G. Luppino,

J. Annis, D. Clowe, J. Huchra, O. Le Fèvre , F. Hammer, E. Shaya, H. Böhringer, U. Briel. I am mostly grateful to the EMSS original group team (in particular T. Maccacaro, J. Stocke, A. Wolter, S. Morris, R. Schild) without whose dedication and persistence in identifying all the EMSS sources, these subsequent achievements would not have been possible. This work received partial financial support from the following grants: NSF grants AST–91199216 and AST–9020680; NASA grants NAG5-2594, NAG5-2914 and NASA-StScI GO-5402.01-93A. This work was partially based on observations with the NSA/ESA *Hubble Space Telescope* obtained at the STScI which is operated by AURA, Inc., under NASA contract NS5-26555.